\documentclass[%
 reprint,
superscriptaddress,
 aps,
]{revtex4-1}

\usepackage{graphicx, amsmath, amssymb, color, upgreek, textcomp}

\begin{document}

\title{Microwave magnon damping in YIG films at millikelvin temperatures}
\author{S. Kosen} \email{sandoko.kosen@physics.ox.ac.uk}
\affiliation{Clarendon Laboratory, Department of Physics, University of Oxford, Oxford OX1 3PU, United Kingdom}
\author{A.\,F. van Loo}
\affiliation{Clarendon Laboratory, Department of Physics, University of Oxford, Oxford OX1 3PU, United Kingdom}
\affiliation{Center for Emergent Matter Science, RIKEN, Wako, Saitama 351-0198, Japan}
\author{D.\,A. Bozhko}
\affiliation{Fachbereich Physik and Landesforschungszentrum OPTIMAS, Technische Universitaet Kaiserslautern, 67663 Kaiserslautern, Germany}
\affiliation{School of Engineering, University of Glasgow,  Glasgow G12 8LT, United Kingdom}
\affiliation{Department of Physics and Energy Science, University of Colorado at Colorado Springs, Colorado Springs CO 80918, USA}
\author{L.\, Mihalceanu}
\affiliation{Fachbereich Physik and Landesforschungszentrum OPTIMAS, Technische Universitaet Kaiserslautern, 67663 Kaiserslautern, Germany}
\author{A.\,D. Karenowska}
\affiliation{Clarendon Laboratory, Department of Physics, University of Oxford, Oxford OX1 3PU, United Kingdom}
\date{\today}

\begin{abstract}
Magnon systems used in quantum devices require low damping if coherence is to be maintained. The ferrimagnetic electrical insulator yttrium iron garnet (YIG) has low magnon damping at room temperature and is a strong candidate to host microwave magnon excitations in future quantum devices. Monocrystalline YIG films are typically grown on gadolinium gallium garnet (GGG) substrates. In this work, comparative experiments made on YIG waveguides with and without GGG substrates indicate that the material plays a significant role in increasing the damping at low temperatures. Measurements reveal that damping due to temperature-peak processes is dominant above 1\,K. Damping behaviour that we show can be attributed to coupling to two-level fluctuators (TLFs) is observed below 1\,K. Upon saturating the TLFs in the substrate-free YIG at 20\,mK, linewidths of $\sim1.4$\,MHz are achievable: lower than those measured at room temperature. 
\end{abstract}

\maketitle

Microwave magnonic systems have been subject to extensive experimental studies for decades.  This work is motivated not only by an interest in their rich basic physics, but also by their potential application as information carriers in beyond-CMOS electronics \cite{Chumak2015, Chumak2017}. Recently, enthusiasm has grown for the study of magnon dynamics at millikelvin (mK) temperatures, the temperature regime in which solid-state microwave quantum systems operate \cite{Huebl2013,Tabuchi2014,Tabuchi2015,Zhang2015, Morris2017, Lachance-Quirion2017,VanLoo2018, Boventer2018,Kosen2018,Goryachev2018}. This work offers the possibility to explore the dynamics of microwave magnons in the quantum regime and to study novel quantum devices with magnonic components \cite{Tabuchi2016,Rusconi2019,Kostylev2019}.

Arguably the most important material in the context of room-temperature experimental magnon dynamics is the ferrimagnetic insulator yttrium iron garnet (Y$_{3}$Fe$_{5}$O$_{12}$, YIG). Pure monocrystalline YIG has the lowest magnon damping of any known material at room temperature \cite{Serga2010} and is produced in the form of bulk crystals and films. Films suitable for use as waveguides in conjunction with micron-scale antennas are grown by liquid-phase epitaxy to a thickness of between 1 and 10$\,\upmu\mathrm{m}$ on gadolinium gallium garnet (Gd$_3$Ga$_5$O$_{12}$, GGG) substrates. The use of GGG is motivated by the need for tight lattice matching to assure a high crystal quality. Recently, YIG films were recognised as promising media for the study of magnon Bose-Einstein condensation and related macroscopic quantum transport phenomena \cite{Demokritov2006,Tiberkevich2019,Bozhko2019,Bozhko2016}. In the context of quantum measurements and information processing, YIG films hold noteworthy promise, however, if they are to be practical, they must be shown to exhibit the same (or better) dissipative properties at mK temperatures as they do at room temperature. 


\begin{figure}
\includegraphics{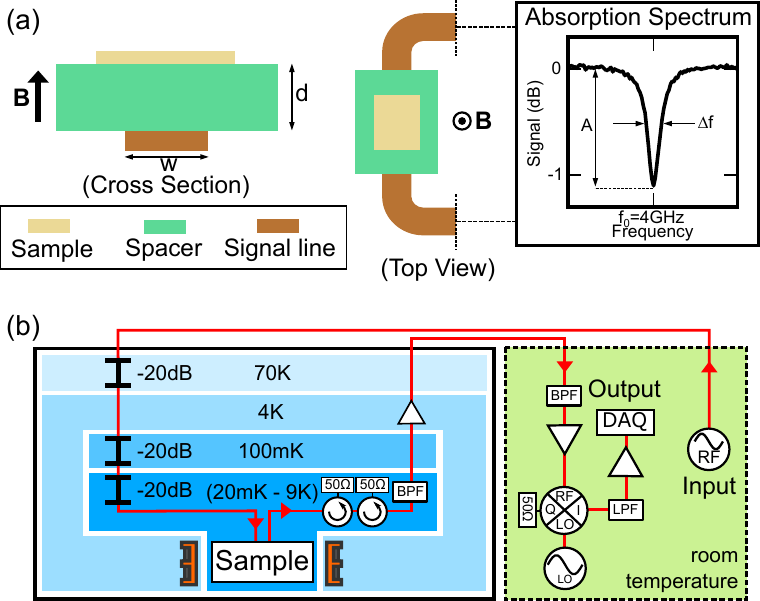}
\caption{(a) The measurement configuration used to characterise the sample's damping. The sample and the microstrip (signal line) are separated from each other by a spacer. When the microwave drive is resonant with the magnons in the sample, a decrease in the transmitted signal is observed. (b) The low-temperature setup and its corresponding data acquisition system at room temperature. \label{figure1}}
\end{figure}

Magnon linewidths in YIG at mK temperatures have thus far only been characterised in \textit{bulk} YIG resonators (specifically, spheres) \cite{Tabuchi2014,Zhang2015,Morris2017,Boventer2018,Pfirrmann2019}. Bulk YIG has been shown to retain its low magnon damping at mK temperatures. However, in the case of YIG \textit{films} grown on GGG, the story is more complex. GGG is a geometrically frustrated magnetic system \cite{Petrenko1999} and it has long been known that at temperatures below $70\,$K, it exhibits paramagnetic behaviour that has been reported to increase damping in films grown on its surface \cite{Danilov1989, Danilov2002, Mihalceanu2018}. The behaviour of GGG at mK temperatures is yet to be thoroughly characterised \cite{Schiffer1994,Schiffer1995, Rousseau2017}, but recent results at mK temperatures have suggested that magnon damping in YIG films grown on GGG is higher than expected if the properties of the YIG system alone are considered \cite{Karenowska2015,Kosen2018,VanLoo2018}. In this work we report a comparative set of experiments on YIG films with and without GGG and move toward a more complete understanding of the damping mechanisms involved. 

We present data from the measurement of two YIG samples:  a $11\,\upmu\mathrm{m}$-thick film and a substrate-free $30\,\upmu\mathrm{m}$-thick film. Both samples are grown using liquid phase epitaxy  with the surface normal of the YIG film (and the substrate) parallel to the $\langle111\rangle$ crystallographic direction. The substrate-free YIG is obtained by mechanically polishing off the GGG until a $30\,\upmu\mathrm{m}$-thick pure YIG film is obtained \cite{Mihalceanu2018}. The corresponding lateral size of each sample can be found in table \ref{table1}. 

\begin{figure}
\includegraphics{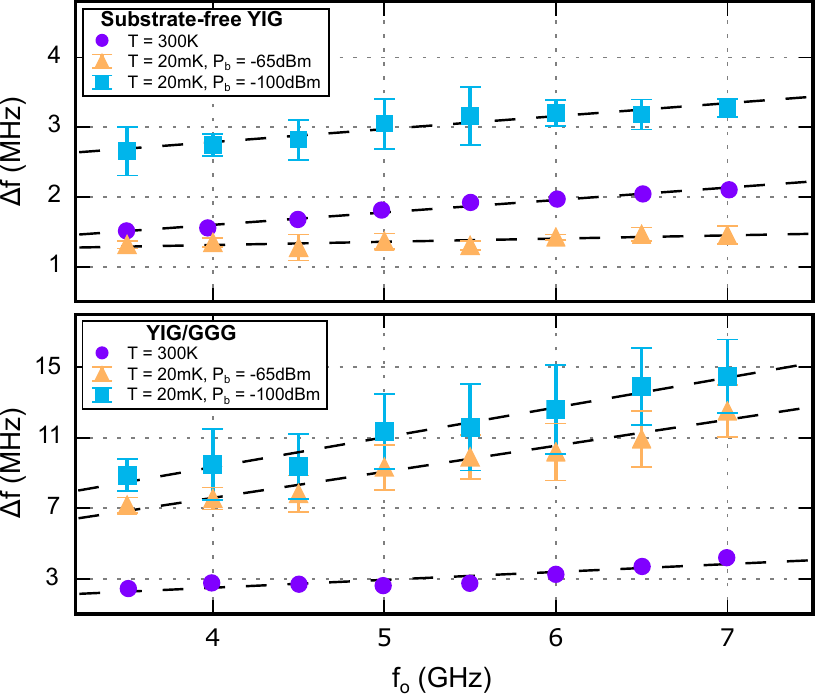}
\caption{Magnon linewidths ($\Delta f$) versus resonance frequency ($f_{\circ}$) for a YIG/GGG film and a substrate-free YIG film. Datasets at room temperature (300\,K, $\bullet$) are obtained with an input power of -25\,dBm. Datasets at 20\,mK are obtained for two input powers $P_\mathrm{b}=-65\,$dBm ($\blacktriangle$), and $P_\mathrm{c}=-100$\,dBm ($\blacksquare$). Each error bar represents the standard deviation of the linewidth values obtained from repeated measurements. Dashed lines are linear fits, the details of these fits are summarised in table \ref{table1}. Note the different scaling of the vertical axes of the plots. \label{compare_roomT_20mK}}
\end{figure}

We measure the damping in both films using the microstrip-based technique illustrated in fig.\,\ref{figure1}(a) \cite{Maksymov2015}. The sample is positioned above a microstrip and magnetised by an out-of-plane magnetic field ($B$). Continuous-wave microwave signals transmitted through the microstrip probe the ferromagnetic resonance of the sample. In the room-temperature experiments, the transmitted signals are measured by connecting the two ends of the microstrip directly to a commercial network analyser. In our low-temperature experiments, the sample is mounted on the mixing-chamber plate of a dilution refrigerator as shown in fig.\,\ref{figure1}(b), similar to that used in Ref.\,\cite{Kosen2018}. A microwave source is used to generate the input microwave signal. At the input line, three 20\,dB attenuators are used to ensure an electrical noise temperature that is comparable to the temperature of the sample. The output signals then pass through two circulators, a bandpass filter, and an amplifier, before they are down-converted to a 500\,MHz signal at room temperature. A DAQ card then digitises the transmitted signal at a 2.5\,GHz sampling frequency. Signals are usually averaged about 50000 times before being digitally down-converted in order to obtain a signal similar to the one shown in fig.\,\ref{figure1}(a). The magnon linewidth is given by the full-width at half maximum (FWHM) of the Lorentzian fit to the transmitted signal. The measurement frequency range is between 3.5\,GHz and 7\,GHz. The low-frequency limit is imposed by the limited bandwidth of the circulators used for our low-temperature measurement setup and the top of the measurement band is determined by the maximum external field that can be applied by our magnet.

The measured damping comprises contributions from the sample and from radiation damping caused by its interaction with the microstrip. In our experiments, radiation damping originates from eddy currents excited in the microstrip by the magnetic field of the magnons \cite{Schoen2015, Kostylev2016a} and can be decreased by increasing the separation between the sample and the microstrip ($d$) at the expense of reducing the measured absorption signal strength ($A$). There is therefore a tradeoff to be made between being able to measure linewidths very close to the intrinsic linewidth of the sample (thick spacer, negligible radiation damping) and being able to achieve sufficient signal-to-noise (SNR) ratio (thin spacer, non-negligible radiation damping). Table \ref{table1} lists the microstrip-sample spacings ($d$) in our experiments. Since earlier experiments suggested that the YIG/GGG linewidth would be higher at low temperature, the YIG/GGG sample is closer to the microstrip to maintain sufficient SNR.

 \begin{table}[t]
 \caption{Comparing results at 300\,K and at 20\,mK}
 \scriptsize
 \centering
 \begin{tabular}{p{1.7cm} | l |l }
 \hline \hline
 & \textbf{YIG/GGG} & \textbf{Substrate-free YIG} \\ \hline
 \textbf{Size} & 2\,mm$\times$3\,mm$\times$10\,$\mu$m &  $\sim$1\,mm$\times$1\,mm$\times$30\,$\mu$m \\ \hline
 $\mathbf{w/d}$ & 1.7\,mm / $70\,\mu$m & 0.9\,mm / $540\,\mu$m \\ \hline
\textbf{300\,K} & $\alpha_\mathrm{1a}=(22\pm 4)\times 10^{-5}$ & $\alpha_\mathrm{2a}=(8.9\pm 0.5)\times 10^{-5}$ \\
  & $\Delta f_\mathrm{\circ,1a}=(0.7\pm0.4)\,$MHz & $\Delta f_\mathrm{\circ,2a} = (0.9\pm0.1)\,$MHz \\ \hline
  \textbf{20\,mK} & $\alpha_{1b}= (74\pm 5 ) \times 10^{-5}$ & $\alpha_\mathrm{2b} = (2.3 \pm 0.7) \times 10^{-5}$ \\
$P_\mathrm{b}$=-65\,dBm & $\Delta f_\mathrm{\circ,1b}=(1.7\pm0.6)\,$MHz & $\Delta f_\mathrm{\circ,2b}=(1.1\pm0.1)$\,MHz \\ \hline
\textbf{20\,mK} & $\alpha_\mathrm{1c} = (85\pm6) \times 10^{-5}$ & $\alpha_\mathrm{2c} = (9.3\pm1.0)\times 10^{-5}$ \\
$P_\mathrm{c}$=-100\,dBm & $\Delta f_\mathrm{\circ,1c}=(2.6\pm0.6)\,$MHz& \ $\Delta f_\mathrm{\circ,2c}=(2.0\pm0.1)\,$MHz\\ \hline
 \end{tabular}\label{table1}
 \end{table}

Within the YIG film itself, the primary contributions to magnon damping are: intrinsic processes \cite{Kasuya1961,Cherepanov1993}, temperature-peak processes \cite{Sparks1964, Gurevich1996}, two-level fluctuator (TLF) processes \cite{Tabuchi2014} and two-magnon processes \cite{Sparks1964, Gurevich1996}. Intrinsic processes are due to interactions with optical phonons and magnons; they are expected to decrease with reducing temperature. Temperature-peak processes originating from interactions with rare-earth impurities are only significant at low temperature (above 1\,K). TLF processes are due to damping sources that behave as two-level systems; they are dominant below 1\,K. Two-magnon processes have their origins in inhomogeneities in the material; in our experiments, they are minimized by magnetising the sample out of plane \cite{McMichael2004, Landeros2008}.
 
Figure \ref{compare_roomT_20mK} compares the magnon linewidth ($\Delta f$) of each sample at 300\,K (room temperature) and at 20\,mK as a function of the ferromagnetic resonance frequency ($f_{\circ}$). Results at 300\,K are obtained by sweeping the input microwave frequency under constant $B$-field. Results at 20\,mK are obtained by sweeping the $B$-field at constant input microwave frequency. In the latter case, the linewidths are measured in terms of magnetic field ($\Delta B$) and converted to units of frequency ($\Delta f$) via the relation $\Delta f =(\gamma/2\pi) \Delta B$, where $\gamma$ is the gyromagnetic ratio. Note that there is no conversion factor other than $\gamma/2\pi$ that is used to translate the low-temperature field-domain data into the frequency domain. A linear fit to $\Delta f = 2\alpha f_{\circ} + \Delta f_{\circ}$ gives the characteristic Gilbert damping constant $\alpha$ (unitless) and the inhomogeneous broadening contribution $\Delta f_{\circ}$. Table \ref{table1} summarises the results of linear fits to data in fig.\,\ref{compare_roomT_20mK}.

We first compare the results at 300\,K and $20\,$mK obtained at relatively high input drive level ($P_\mathrm{b}=-65$\,dBm). The substrate-free YIG shows a measured linewidth decreasing from the room temperature value to approximately $1.4$\,MHz at 20\,mK. The reduction in damping is as anticipated by existing models that describe the intrinsic damping of YIG \cite{Kasuya1961,Sparks1964, Cherepanov1993}. The radiation damping contribution to the linewidth for the substrate-free YIG is small due to the large spacing from the microstrip ($d=540\,\upmu\mathrm{m}$). 

\begin{figure}
\includegraphics{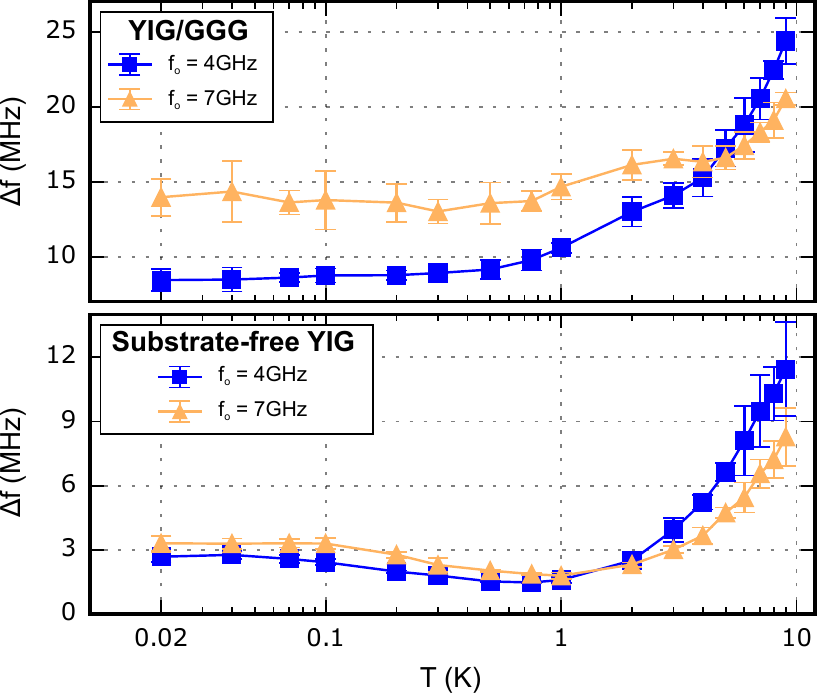}
\caption{Temperature ($T$) dependent magnon linewidths ($\Delta f$) for both YIG/GGG film and substrate-free YIG, measured with input power $P_\mathrm{c}=-100$\,dBm. Note the different scaling of the vertical axes of the plots. \label{temperature_dependent_linewidths}}
\end{figure}

The YIG/GGG sample is substantially closer ($d=70\,\upmu\mathrm{m}$) to the microstrip and its measured $\alpha$ therefore includes a non-negligible radiation damping contribution $\alpha_\mathrm{r}$. In our raw data, uncorrected for this effect, we measure a  damping constant at $20$\,mK ($\alpha_\mathrm{1b}$) that is $3.4$ times larger than the room temperature value ($\alpha_\mathrm{1a}$). Following Ref.\,\cite{Schoen2015}, the radiation damping can be modelled with an equivalent Gilbert damping constant $\alpha_\mathrm{r}=C_\mathrm{g} M_\mathrm{s}$, where $C_\mathrm{g}$ depends on the geometry of the system and $M_\mathrm{s}$ is the saturation magnetisation of the sample. As both 300\,K and 20\,mK measurements are performed with identical sample geometry, it is reasonable to expect that the change in $\alpha_\mathrm{r}$ as the temperature is lowered is due to the change in $M_\mathrm{s}$. Therefore, the increase in $\alpha_\mathrm{r}$ between 20\,mK and 300\,K is determined by the ratio of the saturation magnetisation, i.e. $M_\mathrm{s}(20$\,mK$)/M_\mathrm{s}(300$\,K$)\approx 1.4$ \cite{Maier-Flaig2017}. The fact that we see a significantly larger damping increase ($\alpha_\mathrm{1b}/\alpha_\mathrm{1a}\approx3.4$) in the YIG/GGG and a decrease ($\alpha_\mathrm{2b}/\alpha_{2a}\approx0.26$) in the substrate-free YIG indicates that the GGG plays an important role in increasing the magnon linewidth of the YIG/GGG sample at 20\,mK. 

The parameters $\alpha$ and $\Delta f_{\circ}$ in both samples increase as the input drive level ($P_\mathrm{c}$) reduces as shown in Table \ref{table1}. This behaviour can be explained by the TLF model upon which we shall elaborate later. 

Figure \ref{temperature_dependent_linewidths} shows the temperature dependence of the magnon linewidths for both samples measured at low input power ($P_\mathrm{c}=-100$\,dBm). For the YIG/GGG results in fig.\,\ref{temperature_dependent_linewidths}, the radiation damping contribution ($\alpha_\mathrm{r}=C_\mathrm{g} M_\mathrm{s}$ \cite{Schoen2015}) across the examined temperature range can be considered to be an approximately constant vertical shift to each dataset. This is due to the small change (less than 0.07\%) in $M_\mathrm{s}$ of YIG between $20$\,mK and $9\,$K \cite{Maier-Flaig2017}. 

Above 1\,K, linewidths of both samples increase as the temperature is increased up to 9\,K. In this temperature range, damping is dominated by temperature-peak processes caused by the presence of rare-earth impurities in the YIG \cite{Spencer1959, Sparks1964, Jermain2017, Maier-Flaig2017, Mihalceanu2018}. When temperature-peak processes are dominant, the linewidth of the sample peaks at a characteristic temperature ($T_\mathrm{ch}$) determined by the damping mechanism and the type of impurity. 

\begin{figure}
\includegraphics{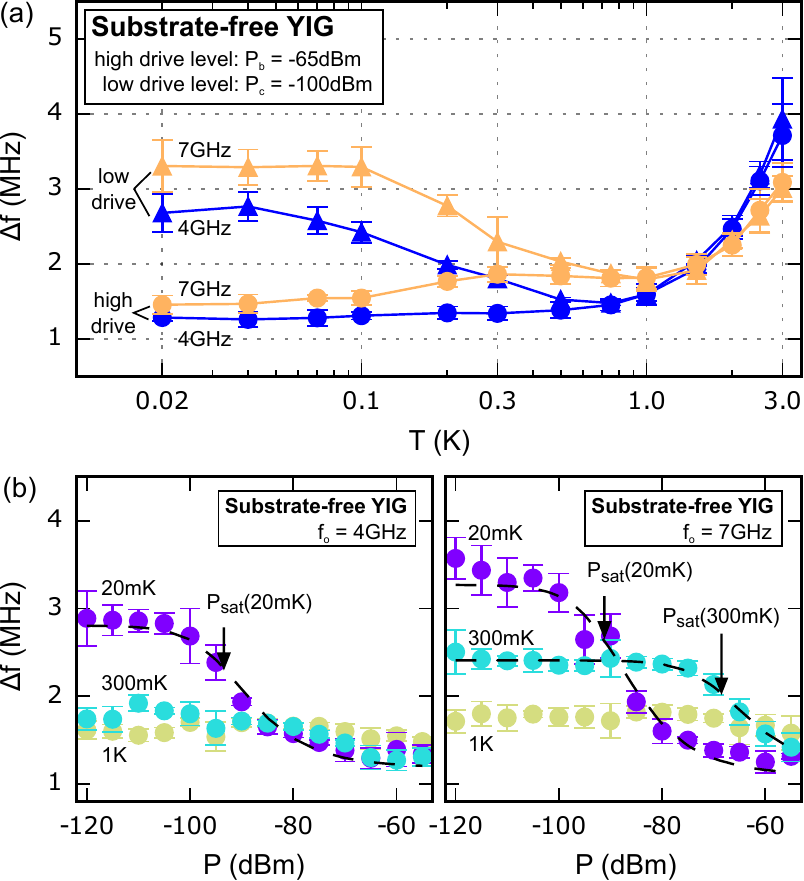}
\caption{Magnon linewidths ($\Delta f$) in the substrate-free YIG film for various temperatures ($T$) and input powers ($P$). (a) Temperature dependent linewidths for two different input powers  $P_\mathrm{b}=-65$\,dBm and $P_\mathrm{c}=-100$\,dBm. (b) Power dependent linewidths at 20\,mK, 300\,mK, and 1\,K. The dashed lines are the fits to the data.}
\label{YIG_powersweep}
\end{figure}

Temperature-peak processes at low temperatures fall into two categories \cite{Sparks1964, Gurevich1996}: those associated with (1) rapidly relaxing impurities, and (2) slowly relaxing impurities. Rapidly relaxing impurities produce a Gilbert-like damping and a characteristic temperature $T_\mathrm{ch}$ that is independent of the magnon resonance frequency $f_{\circ}$. Slowly relaxing impurities produce a non-Gilbert-like damping and a corresponding characteristic temperature that decreases as the resonance frequency $f_{\circ}$ is lowered. The behaviour observed in fig.\,\ref{temperature_dependent_linewidths} at 9\,K, with the linewidth for the $f_{\circ}=4$\,GHz being higher than that at $f_{\circ}=7$\,GHz, indicates that impurities of slowly relaxing type dominate in this temperature range. 

As the temperature is decreased below 1\,K, linewidths for the substrate-free YIG start to increase and eventually saturate as shown in fig.\,\ref{temperature_dependent_linewidths}. This can be explained by the presence of two-level fluctuators (TLFs) and has been previously observed in a bulk YIG \cite{Tabuchi2014}. In the TLF model, the damping sources are modelled as an ensemble of two-level systems with a broad frequency spectrum \cite{Phillips1987, Zmuidzinas2012}. The linewidth contribution can be expressed as
\begin{align}
\Delta f_\mathrm{TLF} = C_\mathrm{TLF}\omega\frac{\tanh{\left(\hbar\omega/2k_\mathrm{B}T\right)}}{\sqrt{1+(P/P_\mathrm{sat})}} \label{eq1}
\end{align}
where $C_\mathrm{TLF}$ is a factor that depends on the TLF and the host material properties. The power-dependent term can be rewritten as $P/P_\mathrm{sat}=\Omega_\mathrm{r}^2 \tau_1\tau_2$, where $\Omega_\mathrm{r}$ is the TLF Rabi frequency, $\tau_1$ and $\tau_2$ are respectively the TLF longitudinal and transverse relaxation times \cite{Gao2008}. 

At high temperatures ($k_\mathrm{B}T \gg hf_\mathrm{TLF} $), thermal phonons saturate the TLFs and therefore the material behaves as if the TLFs were not present. At low temperatures ($k_\mathrm{B}T \ll hf_\mathrm{TLF}$) and low drive levels ($P \ll P_\mathrm{sat}$), most of the TLFs are unexcited. Under these conditions, the TLFs increase the damping of the material by absorbing and re-emitting magnons or microwaves at rates set by their lifetimes, coupling strength, and density. When the drive level is increased past a certain threshold ($P \gg P_\mathrm{sat}$), the TLFs are once again saturated and therefore do not contribute to the damping.

Evidence for the presence of the TLFs is shown in figs.\,\ref{compare_roomT_20mK} and \ref{YIG_powersweep}. The datasets for 20\,mK in fig.\,\ref{compare_roomT_20mK} show that the linewidths for both samples are lower when the drive level is higher ($P_\mathrm{b}$ vs $P_\mathrm{c}$). Figure \ref{YIG_powersweep}(a) shows a similar behaviour in the substrate-free YIG. Above 1\,K, linewidths for both drive levels are similar: an indication that the relevant TLFs are saturated by thermal phonons. The differences in linewidths  for the two drive levels begin to appear as the temperature is lowered below 1\,K. 

Figure \ref{YIG_powersweep}(b) shows the linewidths of the substrate-free YIG as a function of drive level ($P$) at three different temperatures (1\,K, 300\,mK, and 20\,mK). At 1\,K, there is no observable power dependence as the relevant TLFs have been saturated by the thermal phonons. At 20\,mK and 300\,mK, the linewidth increases as the power decreases, saturating at mK temperatures. This is in agreement with the theory previously articulated and the fits shown by dashed lines in fig.\,\ref{YIG_powersweep}(b). The data are fitted using eq.\,(\ref{eq1}) with an additional y-intercept to account for non-TLF linewidth contributions.

For the $f_{\circ}=7$\,GHz dataset in fig.\,\ref{YIG_powersweep}(b), $P_\mathrm{sat}$ at 300\,mK is clearly larger than at 20\,mK. This is in-line with expectations: $\tau_1$ and $\tau_2$ are anticipated to decrease as the temperature is increased, leading to a higher $P_\mathrm{sat}$ (recall that $P_\mathrm{sat}\propto1/\tau_1\tau_2$) \cite{VonSchickfus1977, Golding1981,Gao2008}. The exact temperature dependence of $1/\tau_1\tau_2$ is not clear; in previous experiments, a phenomenological model was suggested with the quantity $1/\tau_1\tau_2$ varying from $T^2$ to $T^4$ \cite{VonSchickfus1977}. This places the ratio $P_\mathrm{sat}(300\,\mathrm{mK})/P_\mathrm{sat}(20\,\mathrm{mK})$ in the range of 23.5\,dB to 47\,dB. The fitted $P_\mathrm{sat}$ values from our data correspond to a ratio of approximately 22.5\,dB, suggestive of a $T^2$ behaviour.

It should be noted that the $f_{\circ}=4$\,GHz, $T=300\,$mK dataset shows a very weak TLF effect since there are sufficient thermal phonons to saturate the TLFs with central frequencies around 4\,GHz; this is not the case for higher frequency datasets taken at the same temperature. A higher $P_\mathrm{sat}$ is also observed at 300\,mK for $f_{\circ}=5\,$GHz and $f_{\circ}=6$\,GHz (data not shown).

Figure \ref{YIG_powersweep}(a) shows that the input power $P_\mathrm{b}=-65$\,dBm used in our experiments is not enough to saturate the relevant TLFs for temperatures between 100\,mK and 1\,K. The datasets obtained with high drive level ($P_\mathrm{b}$) in fig.\,\ref{YIG_powersweep}(a) show that the linewidth difference $\delta f = |\Delta f(f_{\circ}=7\,$GHz$)-\Delta f(f_{\circ}=4\,$GHz$)|$ broadens as the temperature is increased from 100\,mK to 300\,mK, narrowing back as the temperature reaches 1\,K. If a higher drive level is used, $\delta f$ is expected to be smaller at temperatures between 100\,mK and 1\,K.

In conclusion, the substrate GGG on which typical YIG films are grown significantly increases the magnon linewidth at mK temperatures. However, if the substrate is removed, it is possible to obtain YIG linewidths at mK temperatures that are lower than the room-temperature values.  Measured linewidths of both YIG/GGG and substrate-free YIG systems above 1\,K are consistent with the temperature-peak processes typically observed in YIG containing rare earth impurities. Damping due to the presence of unsaturated two-level fluctuators is observed in both YIG/GGG and substrate-free YIG films below 1\,K. We observe the TLF saturation power to be higher at higher temperatures in agreement with the existing literature. We further verify that using high drive level reduces the linewidths of the substrate-free YIG films down to $\sim\,1.4\,$MHz ($f_{\circ}=3.5\,$GHz to 7.0\,GHz) at 20\,mK. 

Looking forward, our measurements suggest that---in the context of the development of magnonic quantum information or measurement systems---it may be worthwhile to investigate the possibility of growing YIG films on substrates other than GGG, or techniques which circumvent the use of a substrate entirely \cite{Levy1999,Balinskiy2017,Delgado2018,Forster2019}. It should be emphasised that the current experimental configuration does not allow us to pinpoint the origin of the TLFs; further investigations into TLFs in YIG would be useful in obtaining high-quality YIG magnonic devices that operate in the quantum regime.

\textit{Note added} - A pre-print by Pfirrmann \textit{et al.} \cite{Pfirrmann2019} recently reported experiments concerning the effect of two-level fluctuators on the linewidth of bulk YIG. This work helpfully complements our investigations into the behaviour of YIG films.

\begin{acknowledgments}
We thank A.A.\,Serga for helpful discussions, and J.F.\,Gregg for the use of his room-temperature magnet. Support from the Engineering and Physical Sciences Research Council grant EP/K032690/1 (SK, AFvL, ADK), the Deutsche Forschungsgemeinschaft Project No. INST 161/544-3 within SFB/TR 49 (DAB, LM), the Indonesia Endowment Fund for Education (SK), and the Alexander von Humboldt Foundation (DAB) is gratefully acknowledged. AFvL is an International Research Fellow at JSPS.
\end{acknowledgments}

\bibliographystyle{apsrev4-1}
\bibliography{linewidth_bib}
\end{document}